\documentclass[aps,prl,amsmath,amssymb,twocolumn,showpacs,superscriptaddress]{revtex4}
\usepackage{graphicx}

\newcommand\textsubscript[1]{\ensuremath{{}_{\text{#1}}}}
\makeatletter
\newcommand\ps@Standard{%
\renewcommand\@oddhead{}%
\renewcommand\@evenhead{}%
\renewcommand\@oddfoot{[Warning: Draw object ignored]}%
\renewcommand\@evenfoot{\@oddfoot}%
\setlength\paperwidth{8.5in}\setlength\paperheight{11in}\setlength\voffset{-1in}\setlength\hoffset{-1in}\setlength\topmargin{1in}\setlength\headheight{12pt}\setlength\headsep{0cm}\setlength\footskip{12pt+0.3701in}\setlength\textheight{11in-1in-0.5909in-0cm-12pt-0.3701in-12pt}\setlength\oddsidemargin{1.248in}\setlength\textwidth{8.5in-1.248in-1.248in}
\renewcommand\thepage{\arabic{page}}
\setlength{\skip\footins}{0.0398in}\renewcommand\footnoterule{\vspace*{-0.0071in}\noindent\textcolor{black}{\rule{0.25\columnwidth}{0.0071in}}\vspace*{0.0398in}}
}
\begin{document}
%\draft
%\twocolumn[
%\hsize\textwidth\columnwidth\hsize\csname @twocolumnfalse\endcsname
%\draft
\title{New insight into cataract formation {--} enhanced
stability through mutual attraction}

\author{A.~Stradner}\affiliation{Physics Department and Fribourg Center for Nanomaterials, University
of Fribourg, CH{}-1700 Fribourg, Switzerland }

\author{G.~Foffi}\email{giuseppe.foffi@epfl.ch}\affiliation{Institut Romand de Recherche Num\'erique en Physique des Mat\'eriaux
(IRRMA), CH{}-1015 Lausanne, Switzerland}\affiliation{Institute of
Theoretical Physics, Ecole Polytechnique F\'ed\'erale de Lausanne
(EPFL), CH{}-1015 Lausanne, Switzerland}

\author{N.~Dorsaz}\affiliation{Institut Romand de Recherche Num\'erique en Physique des Mat\'eriaux
(IRRMA), CH{}-1015 Lausanne, Switzerland}\affiliation{Institute of
Theoretical Physics, Ecole Polytechnique F\'ed\'erale de Lausanne
(EPFL), CH{}-1015 Lausanne, Switzerland}

\author{G.~Thurston}\email{george.thurston@rit.edu}\affiliation{Department of Physics, Rochester Institute of Technology,
Rochester, NY 14623{}-5603, USA }

\author{P.~Schurtenberger}\email{peter.schurtenberger@unifr.ch}\affiliation{Physics Department and Fribourg Center for Nanomaterials, University
of Fribourg, CH{}-1700 Fribourg, Switzerland }

%\date{\today}
\begin{abstract}
  Small-angle neutron scattering experiments and molecular dynamics
  simulations combined with an application of concepts from soft matter
  physics to complex protein mixtures provide new insight into the stability
  of eye lens protein mixtures. Exploring this colloid-protein analogy we
  demonstrate that weak attractions between unlike proteins help to maintain
  lens transparency in an extremely sensitive and non-monotonic manner. These
  results not only represent an important step towards a better understanding
  of protein condensation diseases such as cataract formation, but provide
  general guidelines for tuning the stability of colloid mixtures, a topic
  relevant for soft matter physics and industrial applications.
\end{abstract}

\maketitle

%%%%%%%%%%%%%%%  TEXT  %%%%%%%%%%%%%%%%
It has been recognized for some time that apparently unrelated diseases
including cataract, sickle-cell and Alzheimer's represent a broad class of
'protein condensation diseases' ~\cite{Benedek1997}.  While the study of such
diseases has traditionally focused on detailed properties of the molecules
involved, considerable progress has also been made using statistical and
colloid physics. This approach is based on recognizing that a common feature
of protein condensation diseases is attractive interaction between specific
biological molecules leading to dense phases~\cite{Benedek1997}, which
compromise cell and organ function. A subtle interplay between protein
attractions, repulsions and entropy can lead to these condensed phases, and
the difference between health and disease can hinge on intermolecular
interaction changes as small as thermal energy, $k_{B}T$. The similar
sensitivity of colloidal phase transitions suggests that colloid science tools
can help understand the molecular origins of protein condensation diseases,
and contribute to developing effective preventative measures. In a
complementary fashion, globular proteins have also drawn great attention due
to their suitability as model
colloids~\cite{Piazza2000,Stradner2004,Cardinaux2007}. Proteins have proven
useful for investigating the combined effects of short-ranged attractions and
hard and/or soft repulsions, on phase transitions, aggregation and cluster
formation in a wide range of
colloidal suspensions~\cite{Stradner2004,Cardinaux2007}. \\
Cataract, clouding of the eye lens due to light scattering and a leading cause
of blindness worldwide, is an important protein condensation
disease~\cite{Benedek1997,Benedek1971,Benedek1999,Delaye1983,Bloemendal2004}.
Statistical physics and colloid science have helped rationalize eye lens
protein solution transparency, liquid structure and thermodynamic
properties~\cite{Delaye1983,Bloemendal2004,Siezen1985b,Veretout1989,Schurtenberger1989}.
Eye lens cells contain concentrated solutions of proteins called crystallins.
The three major classes of mammalian crystallins are called ${\alpha}$,
${\beta}$, and ${\gamma}$~\cite{Siezen1985b,Veretout1989}. The lens is
normally highly transparent and refractive, but loss of transparency due to
protein aggregation or phase separation can lead to cataract.  Quantifying and
understanding crystallin interactions and their impact on lens transparency is
therefore a first step towards
possible cataract prevention~\cite{Benedek1999}.\\
The ${\alpha}${}-crystallins are globular, polydisperse, multi{}-subunit, 800
kDa proteins with a diameter of about 18 nm, whose interactions are
well{}-described with a simple hard{}-sphere colloid
model~\cite{Tardieu1998,Finet2001}. The ${\gamma}${}-crystallins are
monomeric, with a molecular weight of 21 kDa and a diameter between 3 and 4 nm
for ${\gamma}$B{}-crystallin. The discovery of a metastable liquid{}-liquid
phase separation provided evidence for short{}-range attractions between
${\gamma}${}-crystallins, and use of the corresponding colloid model has led
to a quantitative description of the phase
behavior~\cite{Schurtenberger1989,Thomson1987,Broide1991,Lomakin1996,Malfois1996,Stradner2005}.\\
%%%%%%%%%%%%%%%%%%%%%%%%%%%%%
\begin{figure}[tbh]
\includegraphics[width=.4\textwidth]{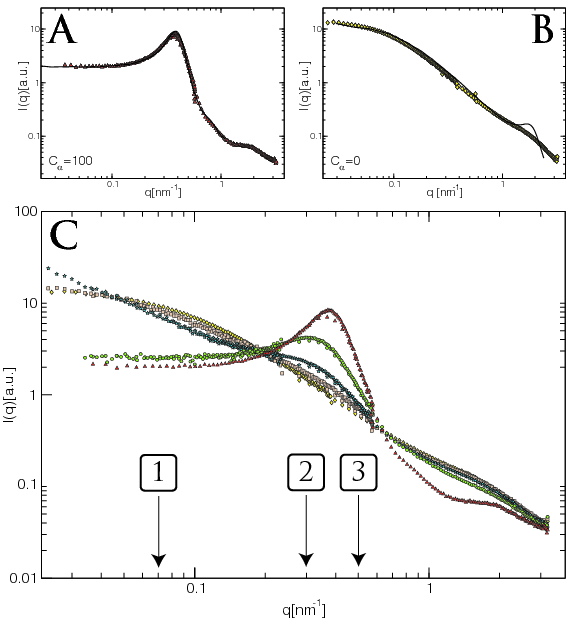}

\caption{(Color online) Scattered intensities $I(q)$ of a $230$ mg/ml
  ($C_\alpha$=100) and a 260 mg/mL ($C_\alpha$=0) solution and their mixtures.
  \textbf{A}, Experimental $I(q)$ of $C_\alpha$=100 (open symbols) together
  with MD computer simulations (full curve). \textbf{B}, Experimental $I(q)$
  of $C_\alpha$=0 (open symbols) together with MD computer simulations (full
  curve). \textbf{C}, $I(q)$ of mixtures containing different amounts of the
  crystallin solutions shown in A and B. Shown are $C_\alpha$=100 (triangles),
  $C_\alpha$=50 (diamonds), $C_\alpha$=25 (stars), $C_\alpha$=12.5 (squares),
  and $C_\alpha$=0 (circles). Numbers indicate $q$-values for composition
  dependence in Fig.~\protect\ref{fig4}.  }
\label{fig1}
\end{figure}
%%%%%%%%%%%%%%%%%%%%%%%%%%%%%
Although there are known synergetic effects in crystallin mixtures associated
with the chaperone activity of ${\alpha}${}-crystallin, which can suppress
induced aggregation of ${\beta}${}- and
${\gamma}${}-crystallin~\cite{Horwitz1992,Putilina2003}, little is known about
the consequences of interactions between unlike crystallins at high
concentrations. Recently, ${\alpha}${}-crystallin has been found to enhance
phase separation of ${\gamma}$B{}-crystallin and lead to partial segregation
of proteins by type in the separated phases~\cite{Thurston2006}.  Therefore we
have started a systematic study of ${\alpha}${}- and ${\gamma}$B{}-crystallin
mixtures up to concentrations found within the eye lens, that combines
small-angle neutron scattering (SANS) and molecular dynamics (MD) simulations
of a coarse-grained colloidal model, appropriate given the hard cores of these
folded, globular proteins. This approach allows for complementary experimental
and computational investigation at the required large
length scales and high number of proteins ($\sim${}30,000 - 60,000).\\
We prepared different mixing ratios from stock solutions of 230 mg/mL
${\alpha}${}-crystallin (denoted as $C_\alpha$=100) and 260 mg/mL
${\gamma}$B{}-crystallin (denoted as $C_\alpha$=0) in 0.1 M sodium phosphate
buffer in D\textsubscript{2}O at a pH of 7.1, with 20mM dithiothreitol to
inhibit protein oxidation and oligomerization~\cite{Thurston2006}. With these
precautions ${\gamma}$B{}-crystallin remained monomeric under all conditions
used. We performed SANS using 1 and 2 mm Hellma quartz cells and varied
wavelengths, sample-to-detector distances and collimation lengths to cover a
\textit{q}{}-range of 0.02{--}3 nm\textsuperscript{{}-1}. All experiments were
performed at a temperature of $25^{\circ}\mathrm{C}$, i.e.
$10^{\circ}\mathrm{C}$ above the critical temperature $T\textsubscript{c}$ of
${\gamma}$B{}-crystallin in the present buffer. The choice of an experimental
temperature close to $T\textsubscript{c}$ of ${\gamma}$B{}-crystallin
amplifies thermodynamic stability variations due to the
presence of alpha crystallin, and thereby enhances sensitivity for discerning effective protein interaction potentials.\\
We established a model for the protein interaction potentials by comparing MD
computer simulations with the experimental scattering intensity $I(q)$ as a
function of the scattering vector $q$.  {Event driven MD
  simulations~\cite{Rapaport1995} were performed in cubic boxes with periodic
  boundary conditions for $N$=32,000 particles for the pure
  ${\alpha}${}-crystallin solution and the mixtures, and $N$=64,000 particles
  for the pure ${\gamma}${}-crystallin solution. The scattering intensities
  were calculated from the partial structure factors $S\textsubscript{ij}(q)$
  (i,j=${\alpha}$,${\gamma}$) obtained from independent MD runs and the
  experimental form factors. In order to account for the experimental
  smearing, we derived a general resolution function and the scattered
  intensities were convoluted before comparison with SANS data~\cite{Dorsaz}.}
From the three different mixing ratios ($C_\alpha$=50, $C_\alpha$=25, and
$C_\alpha$=12, where $C_\alpha$ is defined as 100 x [volume of the
$C_\alpha$=100 solution]/[total volume of the mixture]) investigated,
$C_\alpha$=50 closely resembles the natural ${\alpha}${}- and
${\gamma}${}-crystallin concentrations found in the lens
nucleus~\cite{Siezen1985b}.  Therefore the model development is demonstrated
on this sample, and
its validity is tested by comparison with the remaining samples.\\
First we determined parameters that provide a quantitative description of
$I(q)$ for the individual components. For ${\alpha}${}-crystallin we used a
purely repulsive hard{}-sphere model and a diameter of
$d\textsubscript{${\alpha}$}$ = 17.6 nm~\cite{Finet2001}, which results in
perfect agreement between experimental and simulated $I(q)$ (Fig. 1A). For
$C_\alpha$=0 the strongly enhanced intensity at low $q$ indicates that a
hard{}-sphere model is not sufficient to describe the interactions between
${\gamma}$B{}-crystallins (diameter $d\textsubscript{${\gamma}$}$ = 3.6 nm)
(Fig. 1B). This is a direct consequence of interprotein attraction and
proximity to the critical concentration
\textit{C}\textit{\textsubscript{c}}~\cite{Thomson1987,Broide1991,Malfois1996},
which leads to long{}-wavelength concentration fluctuations that are a
potential source of lens turbidity in cataract~\cite{Benedek1971}.  To account
for this we added a square{}-well attractive potential with a range of 0.25
$d\textsubscript{${\gamma}$}$~\cite{Lomakin1996} and a depth
$u\textsubscript{${\gamma}$${\gamma}$}=1 k\textsubscript{B}T$, where the
temperature $T$ is set to $T=0.7875$ (for $k\textsubscript{B}=1$) for all
simulated mixtures. This temperature $T$ has been fixed to reproduce the
$\gamma$-pure case and results in good agreement between simulated and
experimental $I(q)$ in the low{}-$q$ regime ($q$ $\lesssim$ 0.047 $nm^{-1}$)
relevant for lens transparency (Fig.\ref{fig1}B). It is clear that square well
potentials are rather unphysical in their shape. However, such potentials have
been widely used in studies of colloids with short range potentials due to the
fact that the physics of the system is scarcely dependent on the shape of the
potential when the range is shorter than the diameter of the particle
~\cite{Foffi2002a}. Figure \ref{fig1} indeed demonstrates that simple colloid
models are capable of reproducing $I(q)$ for individual ${\alpha}${}- and
${\gamma}$B{}-crystallin
solutions.\\
In a next step we investigated the mixtures. Figure 1C demonstrates the effect
of adding alpha crystallin, from $C_\alpha$=0 to $C_\alpha$=100.  The forward
scattering first increases and reaches a maximum for $C_\alpha$=25. For
$C_\alpha$=50 it becomes highly suppressed, similar to the situation for pure
{${\alpha}$}{}-crystallin.  In a first attempt to simulate the mixtures we
assumed the interactions between {${\alpha}$}{}- and
{${\gamma}$}{B{}-crystallin to be hard{}-sphere like. The striking
  disagreement between the forward scattering of the corresponding simulation
  with experimental data for $C_\alpha$=50 is shown in Fig.  2. The enormous
  increase of the simulated intensity at low $q$ arises most likely from
  additional depletion{}-induced attractions known to exist in mixtures of
  hard spheres with different sizes~\cite{Mendez-Alcaraz2000}.
%%%%%%%%%%%%%%%%%%%%%%%%%%%%%%
\begin{figure}[tbh]
\includegraphics[width=.43\textwidth]{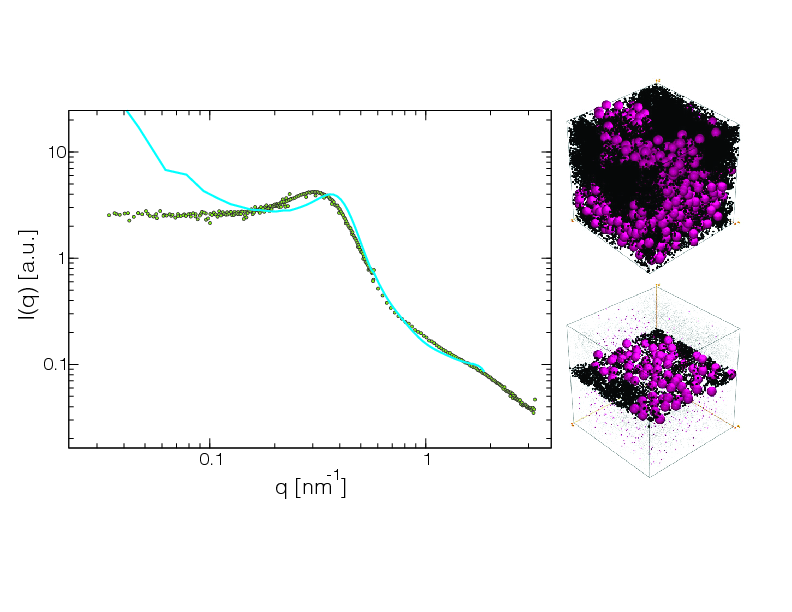}
\caption{(Color online) 
 Scattered intensity $I(q)$ of a mixture of ${\alpha}${}-
and ${\gamma}$B{}-crystallin ($C_{\alpha}=50$) and MD
computer simulation assuming mutual repulsion between unlike proteins.
Left, $I(q)$ of {$C_{\alpha}$}=50 (open symbols)
together with MD simulation of the system. Right, Snapshot from MD
simulation shown on the left. Assuming repulsive hard sphere
interactions only between ${\alpha}${}- (grey spheres) and
${\gamma}$B{}-crystallins (black dots) leads to a strong segregation by
protein type and the corresponding large density fluctuations would
lead to a loss of transparency for visible light. Also shown is a slab of the box for this snapshot.
}
\label{fig2}
\end{figure}
%%%%%%%%%%%%%%%%%%%%%%%%%%%%%%
Such an effect would increase $T\textsubscript{c}$ for liquid{}-liquid phase
separation and result in segregation of the proteins into large domains of
${\alpha}${}-crystallin{}-rich and ${\gamma}$B{}-crystallin{}-rich regions
(Fig. 2).  As a consequence, light scattering would have increased and
transparency would have been lost, in contrast to the experimental data and
the visual appearance of the sample. Such effects are in fact manifest in
${\alpha}${}-${\gamma}$B mixtures at temperatures lower than those
investigated here~\cite{Thurston2006}.  \ However, the low forward scattering
intensity $I(0)$ of the present SANS data indicates that they are suppressed
under the present conditions, closer to body temperature. Nature must have
found ways to circumvent these long{}-wavelength fluctuations and the
accompanying clouding of the eye lens that would have been caused by hard
sphere interactions between ${\alpha}${}- and ${\gamma}$B{}-crystallins, and
we can speculate }that an additional attraction could exist between
these unlike proteins.\\
In a second attempt we thus added an attractive interaction between
${\alpha}${}- and ${\gamma}$B{}-crystallins, assuming that the attractive part
of the potential has the same range as that used for the attraction between
${\gamma}$B{}-crystallins. The depth of this interspecies attractive
potential, $u\textsubscript{${\alpha}$${\gamma}$}$, was chosen to reproduce $I(q)$
and is roughly one half of the ${\gamma}$B{}-${\gamma}$B{}-attraction. The simulated $I(q)$ (Fig. 3)
indeed perfectly reproduces the SANS data throughout the entire $q${}-range. Use
of the novel attractive term leads to good agreement between simulations and
SANS for all mixing ratios (Fig. 4). \
%%%%%%%%%%%%%%%%%%%%%%%%%%%
\begin{figure}[tbh]
\includegraphics[width=.43\textwidth]{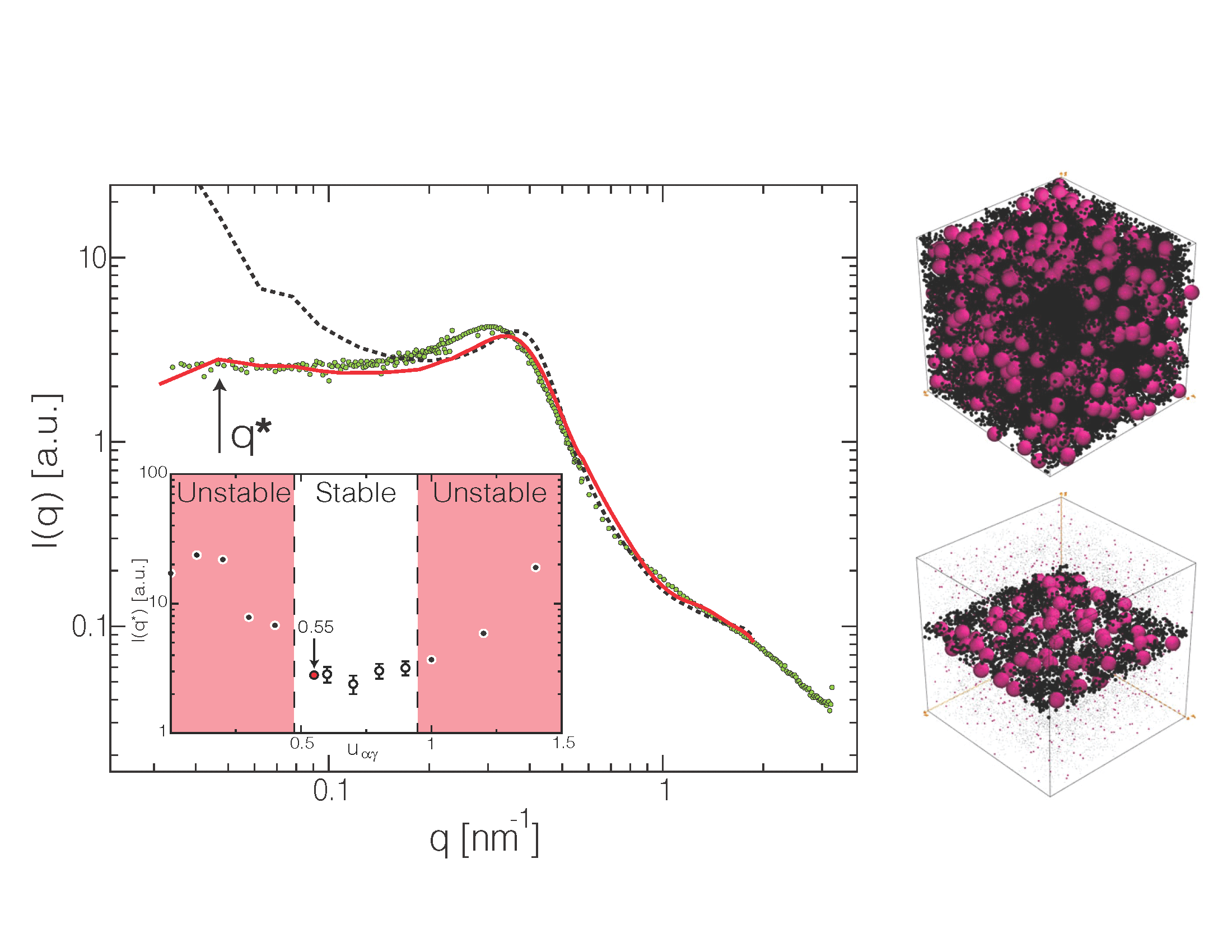}
\caption{(Color online) $I(q)$ of an ${\alpha}${}- and ${\gamma}$B{}-crystallin mixture
  ($C_{\alpha}=50$) and simulations incorporating mutual attraction.
  Left, SANS data from Fig. 2 (open symbols) and simulations without
  (dashed curve) and with added short{}-range attractions (full curve;
  $u_{\alpha\gamma} = 0.55 k_BT$) between
  unlike proteins. Inset: $I(q^* = 0.0467$ nm\textsuperscript{{}-1}) from
  simulations as a function of {$u_{\alpha\gamma}$}. The mixtures become
  unstable for {$u_{\alpha\gamma}\lesssim 0.5 k_BT$} and $\gtrsim 1k_BT$.
  Right, Snapshot and slab representation of the box for this situation as in Fig. 2. The solution is homogeneous and does not give rise to
  increased light scattering. }
\label{fig3}
\end{figure}
%%%%%%%%%%%%%%%%%%%%%%%%%%%
A snapshot of the $C_\alpha$=50 simulation (Fig. 3) suggests a qualitative
explanation: attractions between unlike proteins counterbalance and
efficiently suppress segregation of the two proteins into large domains
~\cite{Louis2002}.  Thus transparency of concentrated crystallin mixtures is
maintained by introducing weak, short{}-range attractions between
${\alpha}${}- and ${\gamma}${}-crystallins, which considerably decrease
$T\textsubscript{c}$ and the corresponding critical fluctuations at a
given temperature.\\
We also performed simulations in which the attraction between ${\alpha}$ and
${\gamma}$B was further increased, to above 100\% of the
${\gamma}$B{}-${\gamma}$B{}-attraction. \ Such stronger attractions again
resulted in enhanced instability (inset of Fig. 3). \ Thus the stability of
these high concentration crystallin mixtures depends on
${\alpha}${}-${\gamma}$B{}-attraction in a manner that is both extremely
sensitive and non{}-monotonic. \ Such non{}-monotonic effects
are a common feature of ternary liquid mixture phase separation ~\cite{Meijering1951, Thurston2006}.\\
There are studies suggesting that ${\gamma}${}-crystallins inhibit the
age{}-related aggregation of ${\alpha}${}-crystallins~\cite{Mach1990}. On the
other hand, numerous investigations have suggested that inhibition of age{}-
or heat{}-induced aggregation of ${\gamma}${}- and ${\beta}${}-crystallins is
associated with the chaperone activity of
${\alpha}${}-crystallins~\cite{Horwitz1992,Putilina2003,Pigaga2006}.
%%%%%%%%%%%%%%%%%%%%%%%%%%%%%%%%%%
\begin{figure}[tbh]
\includegraphics[width=.39\textwidth]{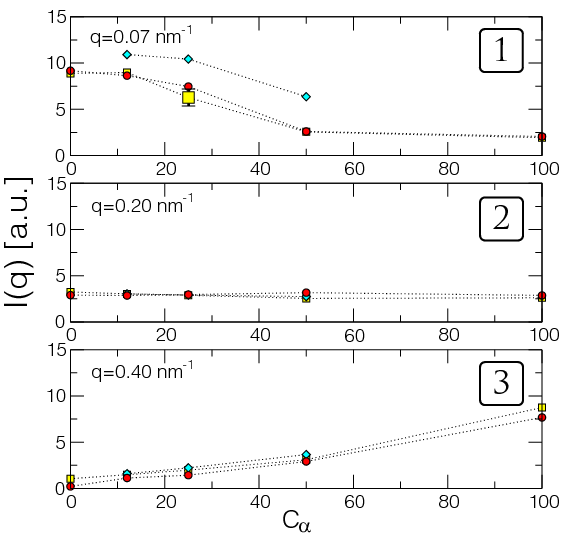}
\caption{(Color online) Comparison
  of $I(q)$ obtained from simulations and SANS as a function of
  ${\alpha}${}-crystallin mixing ratio ({$C_{\alpha}$}) in the
  ${\alpha}${}-${\gamma}$B mixtures.  The predicted low-$q$ portion of $I(q)$ is very sensitive
  to alpha-gamma mutual attraction, unlike the higher-$q$ portions. 
  $I(q)$ at different {$C_{\alpha}$} are  from SANS (circles) and simulations
  without (diamonds) and with (squares) mutual attractions. Shown are data at
  three scattering vectors, $q = 0.07$, $0.2$, and $0.4$ nm$^{-1}$ (the
  numbered labels refer to the arrows in Fig.~\protect\ref{fig1}C ). Unless
  explicitly plotted, error bars are equal or smaller than the symbols.  }
\label{fig4}
\end{figure}
%%%%%%%%%%%%%%%%%%%%%%%%%%%%%%%%%%
  In
contrast to these stabilizing effects of ${\alpha}${}-${\gamma}$ interactions,
it has recently been suggested that the observed increase in the
non{}-covalent association of ${\gamma}${}-crystallins to
${\alpha}${}-crystallins in aging bovine lenses might adversely affect the
optical properties of the aged and/or cataractogenic lens~\cite{Peterson2005},
since strong attractions could lead to larger aggregates and increased
scattering of light. While such a mechanism of increased protein{}-protein
association is a possible route towards cataract~\cite{Peterson2005}, our
experiments and simulations provide evidence that a weak and short{}-range
attraction between ${\alpha}${}- and ${\gamma}$B{}-crystallins can potentially
enhance
transparency in healthy lenses.\\
It is intriguing that a variety of specific molecular mechanisms could
potentially underly this `stabilization due to mutual attraction', since a
simple colloidal ${\alpha}${}-${\gamma}$B attraction is sufficient to
rationalize our observations. Due in part to its small magnitude, the
molecular origin of the inferred, net short{}-range attraction between
${\alpha}${}- and ${\gamma}$B{}-crystallins remains a challenging and open
question, as it is for short{}-range attractions of many globular proteins
including ${\gamma}${}-crystallin and
lysozyme~\cite{Piazza2000,Piazza2004}.\\
It has long been recognized that uniform packing of crystallins at high
concentrations, termed short{}-range order, suppresses index of refraction
fluctuations and reduces light scattering~\cite{Delaye1983}. \ The chemically
specific origins of short{}-range order are key to understanding lens
transparency. Not only do crystallins vary widely in short{}-range order
properties, sensitivity to inter{}-crystallin interactions makes short{}-range
order not a simple combination of individual crystallin properties, as shown
here for ${\alpha}${}- and ${\gamma}$B{}-crystallins. Thus to understand
transparency of crystallin mixtures, high concentration short{}-range order
study, using tools of liquid{}-state and colloid physics, remains a needed
complement to
low concentration investigations.\\
This investigation has provided new insight into the stability and optical
properties of lens protein mixtures that may help understand cataract
formation. The present results also suggest mechanisms to tune the stability
of classical colloidal mixtures, a topic of considerable industrial
importance. This demonstrates the importance of a colloid{}-oriented approach
to proteins as a means of obtaining insight into questions of biological,
medical, as well as fundamental soft
matter physics relevance.
\begin{acknowledgements}
%\acknowledgements{
  This work is based on experiments performed at the Swiss spallation neutron
  source SINQ, Paul Scherrer Institute, Villigen, Switzerland. We are grateful
  for the neutron beam time and we acknowledge the help of our local contact
  J. Kohlbrecher. We thank A. Baldareschi for support and discussions.  This
  work was supported by the Swiss National Science Foundation, the State
  Secretariat for Education and Research (SER) of Switzerland, the Marie Curie
  Network on Dynamical Arrest of Soft Matter and Colloids
  (MCRTN-CT-2003504712), and NIH (U.S.) Grant EY11840.
\end{acknowledgements}

\bibliography{crystallinetal}

\begin{thebibliography}{31}
\expandafter\ifx\csname natexlab\endcsname\relax\def\natexlab#1{#1}\fi
\expandafter\ifx\csname bibnamefont\endcsname\relax
  \def\bibnamefont#1{#1}\fi
\expandafter\ifx\csname bibfnamefont\endcsname\relax
  \def\bibfnamefont#1{#1}\fi
\expandafter\ifx\csname citenamefont\endcsname\relax
  \def\citenamefont#1{#1}\fi
\expandafter\ifx\csname url\endcsname\relax
  \def\url#1{\texttt{#1}}\fi
\expandafter\ifx\csname urlprefix\endcsname\relax\def\urlprefix{URL }\fi
\providecommand{\bibinfo}[2]{#2}
\providecommand{\eprint}[2][]{\url{#2}}

\bibitem[{\citenamefont{Benedek}(1997)}]{Benedek1997}
\bibinfo{author}{\bibfnamefont{G.~B.} \bibnamefont{Benedek}},
  \bibinfo{journal}{Invest Ophthalmol Vis Sci} \textbf{\bibinfo{volume}{38}},
  \bibinfo{pages}{1911} (\bibinfo{year}{1997}).

\bibitem[{\citenamefont{Piazza}(2000)}]{Piazza2000}
\bibinfo{author}{\bibfnamefont{R.}~\bibnamefont{Piazza}},
  \bibinfo{journal}{Curr. Opin. Colloid Int. Sci.}
  \textbf{\bibinfo{volume}{5}}, \bibinfo{pages}{38} (\bibinfo{year}{2000}).

\bibitem[{\citenamefont{Stradner et~al.}(2004)}]{Stradner2004}
\bibinfo{author}{\bibfnamefont{A.}~\bibnamefont{Stradner}}
  \bibnamefont{et~al.}, \bibinfo{journal}{Nature}
  \textbf{\bibinfo{volume}{432}}, \bibinfo{pages}{492} (\bibinfo{year}{2004}).

\bibitem[{\citenamefont{Cardinaux et~al.}(2007, accepted for
  publication)}]{Cardinaux2007}
\bibinfo{author}{\bibfnamefont{F.}~\bibnamefont{Cardinaux}}
  \bibnamefont{et~al.}, \bibinfo{journal}{Phys. Rev. Lett.}
  (\bibinfo{year}{2007, accepted for publication}).

\bibitem[{\citenamefont{Benedek}(1971)}]{Benedek1971}
\bibinfo{author}{\bibfnamefont{G.~B.} \bibnamefont{Benedek}},
  \bibinfo{journal}{Applied Optics} \textbf{\bibinfo{volume}{10}},
  \bibinfo{pages}{459} (\bibinfo{year}{1971}).

\bibitem[{\citenamefont{Benedek et~al.}(1999)}]{Benedek1999}
\bibinfo{author}{\bibfnamefont{G.~B.} \bibnamefont{Benedek}}
  \bibnamefont{et~al.}, \bibinfo{journal}{Prog. Retin. Eye. Res.}
  \textbf{\bibinfo{volume}{18}}, \bibinfo{pages}{391} (\bibinfo{year}{1999}).

\bibitem[{\citenamefont{Delaye and Tardieu}(1983)}]{Delaye1983}
\bibinfo{author}{\bibfnamefont{M.}~\bibnamefont{Delaye}} \bibnamefont{and}
  \bibinfo{author}{\bibfnamefont{A.}~\bibnamefont{Tardieu}},
  \bibinfo{journal}{Nature} \textbf{\bibinfo{volume}{302}},
  \bibinfo{pages}{415} (\bibinfo{year}{1983}).

\bibitem[{\citenamefont{Bloemendal et~al.}(2004)}]{Bloemendal2004}
\bibinfo{author}{\bibfnamefont{H.}~\bibnamefont{Bloemendal}}
  \bibnamefont{et~al.}, \bibinfo{journal}{Prog. Biophys. Mol. Biol.}
  \textbf{\bibinfo{volume}{86}}, \bibinfo{pages}{407} (\bibinfo{year}{2004}).

\bibitem[{\citenamefont{Siezen et~al.}(1985)}]{Siezen1985b}
\bibinfo{author}{\bibfnamefont{R.~J.} \bibnamefont{Siezen}}
  \bibnamefont{et~al.}, \bibinfo{journal}{Proc Natl Acad Sci U S A}
  \textbf{\bibinfo{volume}{82}}, \bibinfo{pages}{1701} (\bibinfo{year}{1985}).

\bibitem[{\citenamefont{V\'er\'etout et~al.}(1989)\citenamefont{V\'er\'etout,
  Delaye, and Tardieu}}]{Veretout1989}
\bibinfo{author}{\bibfnamefont{F.}~\bibnamefont{V\'er\'etout}},
  \bibinfo{author}{\bibfnamefont{M.}~\bibnamefont{Delaye}}, \bibnamefont{and}
  \bibinfo{author}{\bibfnamefont{A.}~\bibnamefont{Tardieu}},
  \bibinfo{journal}{J. Mol. Biol.} \textbf{\bibinfo{volume}{205}},
  \bibinfo{pages}{713} (\bibinfo{year}{1989}).

\bibitem[{\citenamefont{Schurtenberger et~al.}(1989)}]{Schurtenberger1989}
\bibinfo{author}{\bibnamefont{Schurtenberger}} \bibnamefont{et~al.},
  \bibinfo{journal}{Phys. Rev. Lett.} \textbf{\bibinfo{volume}{63}},
  \bibinfo{pages}{2064} (\bibinfo{year}{1989}).

\bibitem[{\citenamefont{Tardieu}(1998)}]{Tardieu1998}
\bibinfo{author}{\bibfnamefont{A.}~\bibnamefont{Tardieu}},
  \bibinfo{journal}{Int J Biol Macromol} \textbf{\bibinfo{volume}{22}},
  \bibinfo{pages}{211} (\bibinfo{year}{1998}).

\bibitem[{\citenamefont{Finet and Tardieu}(2001)}]{Finet2001}
\bibinfo{author}{\bibfnamefont{S.}~\bibnamefont{Finet}} \bibnamefont{and}
  \bibinfo{author}{\bibfnamefont{A.}~\bibnamefont{Tardieu}},
  \bibinfo{journal}{J. of Cryst. Growth} \textbf{\bibinfo{volume}{232}},
  \bibinfo{pages}{40} (\bibinfo{year}{2001}).

\bibitem[{\citenamefont{Thomson et~al.}(1987)}]{Thomson1987}
\bibinfo{author}{\bibfnamefont{J.~A.} \bibnamefont{Thomson}}
  \bibnamefont{et~al.}, \bibinfo{journal}{Proc Natl Acad Sci U S A}
  \textbf{\bibinfo{volume}{84}}, \bibinfo{pages}{7079} (\bibinfo{year}{1987}).

\bibitem[{\citenamefont{Broide et~al.}(1991)}]{Broide1991}
\bibinfo{author}{\bibfnamefont{M.~L.} \bibnamefont{Broide}}
  \bibnamefont{et~al.}, \bibinfo{journal}{Proc Natl Acad Sci U S A}
  \textbf{\bibinfo{volume}{88}}, \bibinfo{pages}{5660} (\bibinfo{year}{1991}).

\bibitem[{\citenamefont{Lomakin et~al.}(1996)\citenamefont{Lomakin, Asherie,
  and Benedek}}]{Lomakin1996}
\bibinfo{author}{\bibfnamefont{A.}~\bibnamefont{Lomakin}},
  \bibinfo{author}{\bibfnamefont{N.}~\bibnamefont{Asherie}}, \bibnamefont{and}
  \bibinfo{author}{\bibfnamefont{G.~B.} \bibnamefont{Benedek}},
  \bibinfo{journal}{J. Chem. Phys.} \textbf{\bibinfo{volume}{104}},
  \bibinfo{pages}{1646} (\bibinfo{year}{1996}).

\bibitem[{\citenamefont{Malfois et~al.}(1996)}]{Malfois1996}
\bibinfo{author}{\bibfnamefont{M.}~\bibnamefont{Malfois}} \bibnamefont{et~al.},
  \bibinfo{journal}{J. Chem. Phys.} \textbf{\bibinfo{volume}{105}},
  \bibinfo{pages}{3290} (\bibinfo{year}{1996}).

\bibitem[{\citenamefont{Stradner et~al.}(2005)\citenamefont{Stradner, Thurston,
  and Schurtenberger}}]{Stradner2005}
\bibinfo{author}{\bibfnamefont{A.}~\bibnamefont{Stradner}},
  \bibinfo{author}{\bibfnamefont{G.~M.} \bibnamefont{Thurston}},
  \bibnamefont{and}
  \bibinfo{author}{\bibfnamefont{P.}~\bibnamefont{Schurtenberger}},
 \bibinfo{journal}{J. Phys.: Condens. Matter}
  \textbf{\bibinfo{volume}{17}}, \bibinfo{pages}{S2805} (\bibinfo{year}{2005}). 

\bibitem[{\citenamefont{Horwitz}(1992)}]{Horwitz1992}
\bibinfo{author}{\bibfnamefont{J.}~\bibnamefont{Horwitz}},
  \bibinfo{journal}{Proc Natl Acad Sci U S A} \textbf{\bibinfo{volume}{89}},
  \bibinfo{pages}{10449} (\bibinfo{year}{1992}).

\bibitem[{\citenamefont{Putilina et~al.}(2003)}]{Putilina2003}
\bibinfo{author}{\bibfnamefont{T.}~\bibnamefont{Putilina}}
  \bibnamefont{et~al.}, \bibinfo{journal}{J. Biol. Chem.}
  \textbf{\bibinfo{volume}{278}}, \bibinfo{pages}{13747}
  (\bibinfo{year}{2003}).

\bibitem[{\citenamefont{Thurston}(2006)}]{Thurston2006}
\bibinfo{author}{\bibfnamefont{G.~M.} \bibnamefont{Thurston}},
  \bibinfo{journal}{J. Chem. Phys.} \textbf{\bibinfo{volume}{124}},
  \bibinfo{pages}{134909} (\bibinfo{year}{2006}).

\bibitem[{\citenamefont{Rapaport}(1995)}]{Rapaport1995}
\bibinfo{author}{\bibfnamefont{D.~C.} \bibnamefont{Rapaport}},
  \emph{\bibinfo{title}{The Art of Molecular Dynamic Simulation}}
  (\bibinfo{publisher}{Cambridge University Press}, \bibinfo{year}{1995}).

\bibitem[{\citenamefont{Dorsaz et~al.}()}]{Dorsaz}
\bibinfo{author}{\bibfnamefont{N.}~\bibnamefont{Dorsaz}} \bibnamefont{et~al.},
  \bibinfo{note}{to be submitted}.

\bibitem[{\citenamefont{Foffi et~al.}(2002)}]{Foffi2002a}
\bibinfo{author}{\bibfnamefont{G.}~\bibnamefont{Foffi}} \bibnamefont{et~al.},
  \bibinfo{journal}{Phys. Rev. E} \textbf{\bibinfo{volume}{65}},
  \bibinfo{pages}{031407} (\bibinfo{year}{2002}).

\bibitem[{\citenamefont{Mendez-Alcaraz and Klein}(2000)}]{Mendez-Alcaraz2000}
\bibinfo{author}{\bibnamefont{Mendez-Alcaraz}} \bibnamefont{and}
  \bibinfo{author}{\bibfnamefont{R.}~\bibnamefont{Klein}},
  \bibinfo{journal}{Phys. Rev. E} \textbf{\bibinfo{volume}{61}},
  \bibinfo{pages}{4095} (\bibinfo{year}{2000}).

\bibitem[{\citenamefont{Louis et~al.}(2002)}]{Louis2002}
\bibinfo{author}{\bibfnamefont{A.~A.} \bibnamefont{Louis}}
  \bibnamefont{et~al.}, \bibinfo{journal}{Phys. Rev. E}
  \textbf{\bibinfo{volume}{65}}, \bibinfo{pages}{061407}
  (\bibinfo{year}{2002}).

\bibitem[{\citenamefont{Meijering}(1951)}]{Meijering1951}
\bibinfo{author}{\bibfnamefont{J.}~\bibnamefont{Meijering}},
  \bibinfo{journal}{Philips Research Reports} \textbf{\bibinfo{volume}{5}},
  \bibinfo{pages}{333} (\bibinfo{year}{1951}) \bibinfo{note}{and} \textbf{\bibinfo{volume}{6}},
  \bibinfo{pages}{183} (\bibinfo{year}{1951}).

\bibitem[{\citenamefont{Mach et~al.}(1990)}]{Mach1990}
\bibinfo{author}{\bibfnamefont{H.}~\bibnamefont{Mach}} \bibnamefont{et~al.},
  \bibinfo{journal}{J. Biol. Chem.} \textbf{\bibinfo{volume}{265}},
  \bibinfo{pages}{4844} (\bibinfo{year}{1990}).

\bibitem[{\citenamefont{Pigaga and Quinlan}(2006)}]{Pigaga2006}
\bibinfo{author}{\bibfnamefont{V.}~\bibnamefont{Pigaga}} \bibnamefont{and}
  \bibinfo{author}{\bibfnamefont{R.~A.} \bibnamefont{Quinlan}},
  \bibinfo{journal}{Exp. Cell Res.} \textbf{\bibinfo{volume}{312}},
  \bibinfo{pages}{51} (\bibinfo{year}{2006}).

\bibitem[{\citenamefont{Peterson et~al.}(2005)\citenamefont{Peterson, Radke,
  and Takemoto}}]{Peterson2005}
\bibinfo{author}{\bibfnamefont{J.}~\bibnamefont{Peterson}},
  \bibinfo{author}{\bibfnamefont{G.}~\bibnamefont{Radke}}, \bibnamefont{and}
  \bibinfo{author}{\bibfnamefont{L.}~\bibnamefont{Takemoto}},
  \bibinfo{journal}{Exp. Eye. Res.} \textbf{\bibinfo{volume}{81}},
  \bibinfo{pages}{680} (\bibinfo{year}{2005}).

\bibitem[{\citenamefont{Piazza}(2004)}]{Piazza2004}
\bibinfo{author}{\bibfnamefont{R.}~\bibnamefont{Piazza}},
  \bibinfo{journal}{Curr. Opin. Colloid Int. Sci} \textbf{\bibinfo{volume}{8}},
  \bibinfo{pages}{515} (\bibinfo{year}{2004}).

\end{thebibliography}=
\bibliographystyle{apsrev}
%%%%%%%%%%%%%%%  FIGURES ON INPUT FILES  %%%%%%%%%%%%%%%%

%\input{figure.tex}
\end{document}